\documentclass[nofootinbib]{revtex4}

\usepackage{booktabs}
\usepackage{graphicx,float}
\usepackage{amsmath,amssymb,amsfonts}
\usepackage{epsfig,color}
\usepackage[thinlines]{easytable}
\usepackage{pdfpages}
\usepackage{array}
\usepackage{cancel}
\usepackage{mathtools}
\usepackage{accents}
\usepackage{subfigure}
\usepackage{enumitem}
\usepackage[dvipsnames]{xcolor}
\usepackage{hyperref}
\usepackage{verbatim}
\usepackage{multirow}
\usepackage{flushend}
\hypersetup{
	colorlinks=true,
	linkcolor=blue,
	filecolor=magenta,
	citecolor=blue
}

\begin{document}
	\hfill  USTC-ICTS/PCFT-25-26
	
	\title{Cosmological Perturbations of Extended NGR Model with Parity Violation}
	
	\author{Yuxuan Kang}
	\email{yxkang@mail.ustc.edu.cn}
	\author{Mingzhe Li}
	\email{limz@ustc.edu.cn}
	\author{Yeheng Tong}
	\email{yhtong19@mail.ustc.edu.cn}
	\affiliation{Interdisciplinary Center for Theoretical Study, University of Science and Technology of China, Hefei, Anhui 230026, China}
	\affiliation{Peng Huanwu Center for Fundamental Theory, Hefei, Anhui 230026, China}

\begin{abstract}
	Recently the modified teleparallel gravity models attracted a lot of interests. In this paper we consider more extensions to the New General Relativity (NGR) model with parity violations. This extended NGR model differs from the normal NGR model by the inclusion of additional parity-odd term that is quadratic in the torsion tensor. By investigating its cosmological perturbations of this model, we find that this model can avoid ghost instabilities in certain regions of the parameter space, where the coefficient of the parity-odd term does not vanish.
\end{abstract}

\maketitle

\section{INTRODUCTION}	

Despite its great success,  general relativity (GR) is not yet a complete theory. The problems it encountered, such as the singularities in black holes and the challenges by $\Lambda$CDM model at large scales of the universe, motivated the flourishing of studies on modifications or extensions to GR \cite{Sotiriou:2008rp,Kobayashi:2019hrl,DeFelice:2010aj}.  One way to modified gravity is the metric-affine approach\cite{Hehl:1994ue}, where the metric and the affine connection are supposed to have independent contributions to the gravity. GR itself can be considered as a special case of metric-affine theory, by requiring that both the torsion and the nonmetricity tensors vanish. In addition to GR, there exist other subclasses of metric-affine models, subjecting to some specific constraints. For instances, one may obtain Einstein-Cartan gravity \cite{Hehl:1976kj,Kibble:1961ba} from a general metric-affine theory by requiring the metric compatibility, and Einstein-Weyl gravity by imposing the condition of torsion free \cite{Weyl:1919fi}. Furthermore, the symmetric teleparallel gravity (STG) \cite{Nester:1998mp,Adak:2006rx} can be obtained if both the torsion and the curvature are set to zero, so that the gravity is attributed to the nonmetricity; and the teleparallel gravity (TG) \cite{Aldrovandi:2013wha,Krssak:2018ywd,Bahamonde:2021gfp} can be got if both the curvature and nonmetricity are set to zero, so that the gravity is attributed to the torsion. In this paper, we will focus on the TG models.

The Lagrangian density of TG model is generally constructed by three simplest scalars expressed in terms of torsion tensor:
\begin{equation}
	\mathcal{I}_1=T^{\mu\nu\rho}T_{\mu\nu\rho},\quad\mathcal{I}_2=T^{\mu\nu\rho}T_{\rho\mu\nu},\quad\mathcal{I}_3=T^{\rho\nu}_{\phantom{12}\rho}T^{\sigma}_{\phantom{1}\nu\sigma}.
\end{equation}
Within the framework of TG theory, a particular instance exists in which the Lagrangian density, denoted by $\mathbb{T}$, formulated as a specific linear combination of $\mathcal{I}_1$, $\mathcal{I}_2$ and $\mathcal{I}_3$. This construction possess the maximum number of gauge symmetries, which is known as the Teleparallel Equivalent of General Relativity (TEGR), exhibits a distinction from GR only by a surface term, hence the name. Beyond TEGR, there are numerous modified TG, which is more complicated. These include, but are not limited to, $f(\mathbb{T})$ gravity \cite{Ferraro:2006jd,Bengochea:2008gz}, $f(\mathbb{T},B)$ gravity \cite{Bahamonde:2015zma}, Teleparallel Horndeski \cite{Bahamonde:2019shr} and others. The significance of TG and modified TG theories lies in their applications to cosmology \cite{Cai:2015emx,Bahamonde:2016cul} and astrophysics \cite{Boehmer:2011gw}, showing their importance in these fields.

One particularly natural theory among the modified TG models is the New General Relativity (NGR) \cite{Hayashi:1979qx}, whose Lagrangian density takes the most general form as a linear combination of $\mathcal{I}_1$, $\mathcal{I}_2$ and $\mathcal{I}_3$ with arbitrary coefficients. The primary constraints of the NGR model have been studied in \cite{Blixt:2018znp,Guzman:2020kgh}, using Hamiltonian analysis, which has determined that the NGR model can be divided into nine different classes according to the primary constraints. Subsequent analyses have examined the dynamics of the nine classes of NGR model \cite{Golovnev:2023ddv,Bahamonde:2024zkb}. The authors of \cite{Golovnev:2023ddv} claimed that the generic case, where the dynamical modes can contain tensor, vector and scalar perturbations, could actually be healthy. However, the authors of \cite{Bahamonde:2024zkb} pointed out that this is not the case because the vector sector of the generic case cannot be ghost free.

In recent years, direct detection of gravitational waves (GWs) from the compact binary systems by the LIGO-Virgo-KAGRA Collaboration has ushered in a new era in gravitational physics \cite{LIGOScientific:2016aoc,LIGOScientific:2017vwq}. Stimulated by these discoveries and the developments in the cosmic microwave background radiation (CMB) experiments \cite{Li:2017drr,CMB-S4:2020lpa}, there has been significant interest in investigating possible parity violating (PV) gravity in the literature. Because different mechanisms of parity violations may induce different effects in GW propagations, in this paper we study a model that extends NGR by incorporating additional parity-odd terms into the action. We aim to take a closer look at the stability issues of the extended NGR model by performing a full cosmological perturbation analysis, with the goal of figuring out in what conditions this model can avoid the ghost problem. We will show that tensor, vector and scalar perturbations can all be dynamical modes without ghost instability, even in the presence of the parity-odd terms.

This paper is organized as follows, In Section \ref{II}, we present a brief introduction of TG theory and introduce the extended NGR model with parity-odd terms. The cosmological perturbation analysis of this model is performed in Section \ref{III} and a special case of the model, where the matter content are absent will be study in Section \ref{IV}. Finally the conclusion will be presented in Section \ref{V}.

From now on, we will use the unit $8\pi G=1$ and the signature of metric is $\{-,+,+,+\}$. The  tensor indices are denoted by Greek $\rho,\mu,\nu...=0,1,2,3$ and by Latin $i,j,k...=1,2,3$ when limiting to spatial components. They are lowered  by the spacetime metric $g_{\mu\nu}$ and raised by its inverse $g^{\mu\nu}$. The local space tensor indices are marked by $A,B,C...=0,1,2,3$ and by $a,b,c...=1,2,3$ when referring to spatial components. These are lower and raise by Minkowski metric $\eta_{AB}$ and its inverse $\eta^{AB}$.  In addition, we denote the Levi-Civita connection and its corresponding covariant derivative operator as $\mathring{\Gamma}^\rho_{\phantom{1}\mu\nu}=\frac{1}{2}g^{\rho\lambda}\left(\partial_\mu g_{\lambda\nu}+\partial_\nu g_{\mu\lambda}-\partial_\lambda g_{\mu\nu}\right)$ and $\mathring{\nabla}$, which are separated from the spacetime affine connection $\Gamma^\rho_{\phantom{1}\mu\nu}$ and its associated covariant derivative operator $\nabla$. The Levi-Civita tensor is  $\varepsilon_{\mu\nu\rho\sigma}=\sqrt{-g}\,\epsilon_{\mu\nu\rho\sigma}$, where $\epsilon_{\mu\nu\rho\sigma}$ is a totally antisymmetric symbol with $\epsilon_{0ijk}=\epsilon_{ijk}\equiv\epsilon^{ijk}$ and $\epsilon_{0123}=-\epsilon^{0123}=1$.

\section{The NGR Model With Parity-Odd Terms}\label{II}

In metric-affine theory, both metric $g_{\mu\nu}$ and affine connection $\Gamma^\rho_{\phantom{1}\mu\nu}$ are regarded as fundamental variables and the TG can be considered as a metric-affine theory with two constrains,  which are curvature-free and metric-compatible:
\begin{equation}\label{2}
	R^\rho_{\phantom{1}\sigma\mu\nu}=\partial_\mu\Gamma^\rho_{\phantom{1}\nu\sigma}-\partial_\nu\Gamma^\rho_{\phantom{1}\mu\sigma}+\Gamma^\rho_{\phantom{1}\mu\lambda}\Gamma^\lambda_{\phantom{1}\nu\sigma}-\Gamma^\rho_{\phantom{1}\nu\lambda}\Gamma^\lambda_{\phantom{1}\mu\sigma}=0,\quad\nabla_\rho g_{\mu\nu}=\partial_\rho g_{\mu\nu}-\Gamma^\lambda_{\phantom{1}\rho\mu}g_{\lambda\nu}-\Gamma^\lambda_{\phantom{1}\rho\nu}g_{\mu\lambda}=0,
\end{equation}
where $\nabla_\rho g_{\mu\nu}\equiv Q_{\rho\mu\nu}$ is the nonmetricity tensor. Since both curvature tensor and nonmetricity tensor all vanish in TG theory, the gravity is determined by torsion $T^\rho_{\phantom{1}\mu\nu}=2\Gamma^\rho_{\phantom{1}\left[\mu\nu\right]}$ and the affine connection can be expressed in terms of Levi-Civita connection and torsion tensor: 
\begin{equation}
	\Gamma^\rho_{\phantom{1}\mu\nu}=\mathring{\Gamma}^\rho_{\phantom{1}\mu\nu}+\frac{1}{2}\left(\, T^\rho_{\phantom{1}\mu\nu}-T_{\mu\nu}^{\phantom{12}\rho}-T_{\nu\mu}^{\phantom{12}\rho}\,\right).
\end{equation}
Besides, the TG theory can also be described with the language of tetrad $e^A_{\phantom{1}\mu}$ and spin connection $\omega^A_{\phantom{1}B\mu}$ which relates the metric $g_{\mu\nu}$ and affine connection through the following equations:
\begin{equation}
	g_{\mu\nu}=\eta_{AB}e^A_{\phantom{1}\mu}e^B_{\phantom{1}\nu},\quad \Gamma^\rho_{\phantom{1}\mu\nu}=e_A^{\phantom{1}\rho}\left(\partial_\mu e^A_{\phantom{1}\nu}+\omega^A_{\phantom{1}B\mu}e^B_{\phantom{1}\nu}\right).
\end{equation} 
Then the torsion tensor can be generally expressed as:
\begin{equation}
	T^\rho_{\phantom{1}\mu\nu}=2e_A^{\phantom{1}\rho}\left(\partial_{ \left[ \mu\right.} e^A_{\left. \phantom{1}\nu \right]}+\omega^A_{\phantom{1}B\left[\mu\right.}e^B_{\left.\phantom{1}\nu\right]}\right).
\end{equation}
Furthermore, the constrains required in (\ref{2}) imply that spin connection can be expressed as:
\begin{equation}
	\omega^A_{\phantom{1}B\mu}={\left(\Lambda^{-1}\right)}^A_{\phantom{1}C}\partial_\mu\Lambda^C_{\phantom{1}B},
\end{equation}
where $\Lambda^A_{\phantom{1}B}\in SO(3,1)$ is an arbitrary position dependent Lorentz matrix, satisfying the relation $\eta_{AB}\Lambda^A_{\phantom{1}C}\Lambda^B_{\phantom{1}D}=\eta_{CD}$ at any spacetime point. Thus, the TG theory can be described within the tetrad $e^A_{\phantom{1}\mu}$ and the Lorentz matrix $\Lambda^A_{\phantom{1}B}$.

Generally, the action of NGR model has the following form:
\begin{equation}\label{7}
	S=-\frac{1}{2}\int d^4 x \|e\|\biggl[c_1T_{\rho \mu \nu}T^{\rho \mu \nu}+c_2T_{\rho \mu \nu}T^{\nu \mu \rho}+c_3T_\mu T^\mu\biggr],
\end{equation}
where $T_\mu\equiv T^\sigma_{\phantom{1}\mu\sigma}$ and  $\|e\|=\sqrt{-g}$. If one takes $c_1:c_2:c_3=1:2:-4$, then the action just becomes TEGR. And because of the identity $-\mathring{R}(e)=\mathbb{T}+2\mathring{\nabla}_\mu T^\mu$ we have known \cite{Bahamonde:2021gfp}, the TEGR action is equivalent to Einstein-Hilbert action up to a
surface term with the curvature scalar $\mathring{R}(e)$, which is defined by Levi-Civita connection expressed in terms of metric, can be fully constructed from the tetrad.

In TG gravity, there are totally four scalar invariants which are parity-odd and quadratic with the torsion tensor \cite{Iosifidis:2018zwo,Li:2022mti}:
\begin{equation}
	P_1=\frac{1}{2}\varepsilon^{\mu\nu\rho\sigma}T^\lambda_{\phantom{1}\mu\nu}T_{\lambda\rho\sigma}~,~P_2=\varepsilon^{\mu\nu\rho\sigma}T_\mu T_{\mu\rho\sigma}~,~P_3=\varepsilon^{\mu\nu\rho\sigma}T^\lambda_{\phantom{1}\mu\nu}T_{\rho\sigma\lambda}~,~P_4=\varepsilon^{\mu\nu\rho\sigma}T^{\phantom{12}\lambda}_{\mu\nu}T_{\rho\sigma\lambda}~,
\end{equation}
where $P_1$ are known as Nieh-Yan density which is a total covariant divergence term \cite{Nieh:1981ww}. With a coupling Lagrangian between an 
scalar field and the Nieh-Yan density added to TEGR action, one can construct a parity violating model without ghost instability, which is called Nieh-Yan modified Teleparallel Gravity (NYTG) model \cite{Li:2020xjt,Li:2021wij}. However, the four invariants given above are not independent. It has been shown proved in Ref.\cite{Iosifidis:2018zwo,Li:2022mti} that there are two identities with these parity odd terms:
\begin{equation}
	P_3=-P_1+P_2,\quad P_4=\frac{1}{2}P_1-P_2,
\end{equation}  
leaving only two invariants $P_1$ and $P_2$  independent. 

Now we consider adding these invariants to NGR model. The Nieh-Yan density is just a surface term if there exist no coupling fields. As a result, the action we are concerned in this paper is:
\begin{equation}\label{action}
	\begin{aligned}
		S&=-\frac{1}{2}\int d^4 x \| e\| \mathcal{L}_g +S_m,\\
		\mathcal{L}_g&=c_1T_{\rho \mu \nu}T^{\rho \mu \nu}+c_2T_{\rho \mu \nu}T^{\nu \mu \rho}+c_3T_\mu T^\mu-\frac{1}{2}c_4\varepsilon^{\mu \nu \rho \sigma}T_{\mu}T_{\nu \rho \sigma},
	\end{aligned}
\end{equation}	
where $S_m$ refers to the action of matter that is minimally-coupled. If $c_4=0$, the action (\ref{action}) returns to NGR model. Moreover, this model naturally avoids the Ostrogradski ghost modes, which caused by higher-order derivatives, since the action only contains the first-order derivatives of the basic variables. The variations of the action (\ref{action}) with respect to $e^A_{\phantom{1}\mu}$ and $\Lambda^A_{\phantom{1}B}$ lead to the equations of motion (EOMs):
\begin{align}
	N^{\mu\nu}~&=~T^{\mu\nu},\label{eom}\\
	N^{[\mu\nu]}&=~0\label{11.b},
\end{align}
where $T^{\mu\nu}=(2/\sqrt{-g})(\delta S_m/\delta g_{\mu\nu})$ is the energy-momentum tensor, and
\begin{equation}
	N^{\mu\nu}=-\frac{1}{\| e\|} \partial_\sigma \left(\| e\| S_{\tau}^{\phantom{1} \sigma \nu}\right)g^{\tau\mu}+\Gamma^\tau_{\phantom{1}\gamma\sigma}S_{\tau}^{\phantom{1}\sigma\nu}g^{\gamma\mu}+\frac{1}{2} g^{\mu\nu} \mathcal{L}_g,
\end{equation}
where $S_{\tau}^{\phantom{1} \mu \alpha}$ is the superpotential:
\begin{equation}
	S_{\tau}^{\phantom{1} \mu \alpha}=2\left[ c_1T_\tau ^{\phantom{1} \mu \alpha}+c_2T^{\left[ \mu \phantom{1}\alpha \right]}_{\phantom{12} \tau}-c_3 \delta^{\left[ \mu\right.}_{\tau} T^{\left.\alpha \right] }+\frac{c_4}{4}\varepsilon^{\rho\sigma\mu\nu}\left(\delta^\alpha_\tau T_{\sigma\rho\nu}+\delta^\alpha_\rho g_{\sigma\tau}T_{\nu}\right) \right] .
\end{equation}

The EOM of $\Lambda^A_{\phantom{1}B}$ is just the antisymmetric part of the EOM (\ref{eom}). As has been shown in Ref.\cite{Li:2021wij}, the reason is that the change caused by $\delta\Lambda^A_{\phantom{1}B}$ can be equivalent to the change cause by $\delta e^A_{\mu}$ because of the local Lorentz invariance of the action. The local Lorentz transformation is
\begin{equation}
	e^A_{\phantom{1}\mu}\rightarrow {\left(L^{-1}\right)}^A_{\phantom{1}B}e^B_{\phantom{1}\mu},\quad\Lambda^A_{\phantom{1}B}\rightarrow\Lambda^A_{\phantom{1}C}L^C_{\phantom{1}B},
\end{equation}
where $L^A_{\phantom{1}B}$ is also a Lorentz matrix, and both torsion tensor and metric remain unchanged under such a transformation, leaving the action also invariant. Therefore, the local Lorentz invariance allow us to choose a gauge where $\Lambda^A_{\phantom{1}B}=\delta^A_B$ and $\omega^A_{\phantom{1}B\mu}=0$, which is called Weitzenb\"{o}ck gauge. Once the Weitzenb\"{o}ck gauge is chosen, six degrees of freedom (DoFs) can be gauge away in the gravity sector but the local Lorentz invariance will no longer hold. In the following sections, we will always choose the Weitzenb\"{o}ck gauge.

In addition to the local Lorentz invariance, the model (\ref{action}) is also diffeomorphism invariant, which allow us to gauge away other four DoFs in the gravity sector. However, the model still have twelve DoFs while the GR only have six after taking the gauge. As we can see in the next section, the extra six variables will appear in the quadratic action for perturbations unless the coefficients in (\ref{action}) are chosen as the TEGR case. In the following sections we will choose spatially flat gauge \cite{Riotto:2002yw} during calculations.

\section{Application To Cosmology}\label{III}

\subsection{Background Evolution}

Now we apply our model (\ref{action}) to cosmology. For simplicity, we take the spatially flat Friedmann-Robertson-Walker(FRW) universe.  The background of the tetrad is parameterized as $e^A_{\phantom{1} \mu}=a(\eta)\delta^A_\mu$, so the background line element has the simple form:
\begin{equation}
	{ds}^2=a^2(\eta)\left(-d\eta^2+\delta_{ij}dx^idx^j\right),
\end{equation}
where $a(\eta)$ is the scale factor and $\eta$ is the conformal time. After substituting the metric given above into (\ref{eom}), one can easily find out that the parity odd term in this model have no contributions to the background dynamics. Therefore, the background equations are similar to those usually seen in GR:
\begin{equation}\label{frd}
	3 \mathcal{H}^2 W= a^2\rho,\quad -\left(2\mathcal{H}'+\mathcal{H}^2\right)W=a^2 p,
\end{equation}
with a difference of coefficient
\begin{equation}
	W=-c_1-\frac{c_2}{2}-\frac{3c_3}{2}.
\end{equation}
Here the prime represents for the derivative with respect to the conformal time and $\mathcal{H}=a'/a$ is the conformal Hubble rate. As usual, $\rho$ and $p$ are energy density and pressure of the matter, respectively, and  $W\geq0$ is required since the energy density $\rho$ should not be negative. One can straightly tell that if $c_1=1/4$, $c_2=2c_1$ and $c_3=-4c_1$, which lead to $W=1$, the eq.(\ref{frd}) are just Friedmann equations in GR, which shows the equivalence between GR and TEGR.
 
A degenerate case that $W=0$ leads to a non-trivial gravity evolution in vacuum with vanishing $\rho$ and $p$, which will be discussed later, and we just take $W>0$ in this section.

\subsection{Cosmological Perturbations In New GR}

In this section, we introduce the linear cosmological perturbations of the model. With the Scalar-Vector-Tensor decomposition, the perturbation of tetrad can be written as follows \cite{Izumi:2012qj,Li:2022mti}:
\begin{equation}
	\begin{aligned}
		&e^0_{\phantom{1}0}=a\left(1+A\right),~e^0_{\phantom{1}i}=a\left(\partial_i\beta+\beta_i\right),~e^i_{\phantom{1}0}=a\left(\partial_i\gamma+\gamma_i\right),\\
		&e^i_{\phantom{1}j}=a\left[\left(1-\psi\right)\delta_{ij}+\partial_i\partial_j\alpha+\partial_j\alpha_i+\epsilon_{ijk}\partial_k\lambda+\epsilon_{ijk}\lambda_k+\frac{1}{2}h^T_{ij}\right].
	\end{aligned}
\end{equation} 

Thus the perturbed metric components have the following forms:
\begin{equation}
	\begin{aligned}
		&g_{00}=-a^2\left(1+2A\right),~g_{0i}=a^2\left(\partial_iB+B_i\right),\\
		&g_{ij}=a^2\left[\left(1-2\psi\right)\delta_{ij}+2\partial_i\partial_j\alpha+\partial_i\alpha_j+\partial_j\alpha_i+h^T_{ij}\right],		
	\end{aligned}
\end{equation}
where $B=\gamma-\beta$ and $B_i=\gamma_i-\beta_i$. All the vector perturbations $\alpha_i, \beta_i, \gamma_i$ satisfy the transverse condition, $i.e.$, $\partial_i\alpha_i=\partial_i\beta_i=\partial_i\gamma_i=0$; and the tensor perturbations are transverse and traceless, i.e., $\partial_ih^T_{ij}=\delta^{ij}h^T_{ij}=0$. Even though we have taken the Weitzenb\"{o}ck gauge, the diffeomorphism invariance is still preserved, which allow us to further choose some specific gauge. The diffeomorphism transformation generated by an infinite small vector $\xi^\mu=(\xi^0,\xi_i+\partial_i\xi)$ is \cite{Li:2021wij}
\begin{equation}
	\begin{aligned}
		&A\rightarrow A-\xi^{\prime0}-\mathcal{H}\xi^{0}~,~\psi\rightarrow\psi+\mathcal{H}\xi^{0}~,~\beta\rightarrow\beta-\xi^{0}~,~\gamma\rightarrow\gamma-
		\xi'~,~\lambda\rightarrow\lambda~,\\
		&\alpha\rightarrow\alpha-\xi~,~\beta_{i}\rightarrow\beta_{i}~,~\gamma_{i}\rightarrow\gamma_{i}-\xi_{i}'
		~,~\alpha_{i}\rightarrow\alpha_{i}-\xi_{i}~,~\lambda_{i}\rightarrow\lambda_{i}~,~h^{T}_{ij}\rightarrow h^{T}_{ij}~,
	\end{aligned}
\end{equation} 
where $\xi_i$ also satisfy the transverse condition. In this paper, we will take spatially flat gauge, \textit{i.e.}, $\psi=\alpha=\alpha_i=0$.

\subsection{Quadratic Action For Tensor Perturbations}

After substituting the linear perturbations of tetrad into the model, one can directly obtain the quadratic action for tensor perturbations:
\begin{equation}\label{26}
	\begin{aligned}
		S^{(2)}_T&=\int d\eta d^3x~ \frac{a^2}{8}  \Big[-\left(2c_1+c_2\right)\eta^{\alpha\beta}\partial_\alpha h^{T}_{ij}\partial_\beta h^{T}_{ij}+3c_4\mathcal{H}\epsilon^{i j k}h^{T}_{il}\partial_jh^{T}_{kl}\Big]\\&=\sum_A \int d\eta d^3\vec{k}~ \frac{a^2}{4}\left[ \left(2c_1+c_2\right)h^{\prime2}_A-k^2\left( 2c_1+c_2-\frac{3c_4\mathcal{H}\mathfrak{p}_A}{k}\right) h^2_A\right],
	\end{aligned}
\end{equation}
where we have transformed this action into Fourier space, $A=L,R$ represents the left/right-handed polarization: $\mathfrak{p}_L=-1,\ \mathfrak{p}_R=1$, and we mark $h^{\prime *}_Ah'_A\rightarrow h^{\prime2}_A,\ h^*_Ah_A\rightarrow h^2_A$ for simplicity. The Fourier expansion of $h_{ij}^T$ is defined as follows:
\begin{equation}
	h_{ij}^T(t,\vec{x})=\frac{1}{(2\pi)^{3/2}}\int d^3\vec{k}~h_{ij}^T(t,\vec{k}) \, e^{i\vec{k}\cdot\vec{x}}~,~h_{ij}^T(t,\vec{k})=\sum_{A=L,R}h_A(t,\vec{k}) \, \hat{e}^A_{ij}(\vec{k}),
\end{equation}
where the bases satisfy that $\hat{e}^A_{ij}(\vec{k})\,\hat{e}^{B*}_{ij}(\vec{k})=2\delta^{AB}$ and $\epsilon_{ikl}\hat{k}_l\,\hat{e}^A_{jk}(\vec{k})=\mathrm{i}\mathfrak{p}_A\,\hat{e}^A_{ij}(\vec{k})$. Then it's obvious that the tensor perturbations would be free of ghost instability if and only if 
\begin{equation}\label{tensor}
	2c_1+c_2\geq0.
\end{equation}

The equation for tensor perturbations derived from (\ref{26}) is
\begin{equation}\label{eomh}
	h_A''+2\mathcal{H}h_A'+\omega_A^2h_A=0,
\end{equation}	
where $\omega_A^2=k^2-3\mathfrak{p}_Ac_4\mathcal{H}k/(2c_1+c_2)$.	From (\ref{eomh}), one can find that GWs with different helicities will have different phase velocities: $v_p^A=\omega_A/k\approx1-3\mathfrak{p}_Ac_4\mathcal{H}/\left[2\left(2c_1+c_2\right)k\right]\equiv1+\mathfrak{p}_AaM_{PV}/(2k)$. This is the so-called velocity birefringence phenomenon of GWs. Through the full Bayesian inference on the GW events of binary black hole merges (BBH), an upper bound on the parameters of NYTG model was placed in Ref.\cite{Wu:2021ndf}: $M_{PV}<6.5\times10^{-42}$GeV. In our framework this implies the constraint: $|3c_4\mathcal{H}/\left[a(2c_1+c_2)\right]|<6.5\times10^{-42}$GeV.

\subsection{Quadratic Action For Vector Perturbations}

The quadratic action for vector perturbations in Fourier space is: 
\begin{equation}\label{28}
	\begin{aligned}
		S^{(2)}_v=&-\dfrac{1}{2}\sum_A\int d\eta d^3\vec{k}~ a^2\Big[Z\beta^{\prime2}_A+k^2Z\lambda_A^2+4\mathfrak{p}_AkZ\mathcal{H}\beta_A\lambda_A-2c_1k^2\beta_A^2+2c_2k^2\beta_A\gamma_A\\
		&-2c_1k^2\gamma_A^2+\left(2c_2-4c_1\right)\lambda^{\prime2}_A+\left(2c_2-4c_1\right)\mathfrak{p}_Ak\gamma_A\lambda_A'+\left(4c_2+2c_3\right)\mathfrak{p}_Ak\beta_A\lambda_A'\\
		&+c_4\left(2\beta_A'\lambda_A'-k^2\beta_A\lambda_A-k^2\lambda_A\gamma_A-2\mathfrak{p}_Ak\mathcal{H}\beta_A^2+\mathfrak{p}_Ak\mathcal{H}\lambda_A^2-\mathfrak{p}_Ak\gamma_A\beta_A'\right)\Big].
	\end{aligned}
\end{equation}
Here we define $Z=2c_1+c_2+c_3$ and simply marked $\beta_A\lambda_A^*,~\beta_A^*\lambda_A\rightarrow\beta_A\lambda_A$, and so on. One can see that $\gamma_A$ is not a dynamical field and the variation with respect to it yields the constraint:
\begin{equation}\label{cec10}
	 -4c_1k^2\gamma_A+2c_2k^2\beta_A+2\left(c_2-2c_1\right)\mathfrak{p}_Ak\lambda_A'-c_4k^2\lambda_A-c_4\mathfrak{p}_Ak\beta_A'=0.
\end{equation}
It's clear that $\gamma_A$ is just a Lagrange multiplier if $c_1=0$, while if $c_1\neq0$ we should solve $\gamma_A$ and substitute it back into the quadratic action.\\

\subsubsection{$c_1=0$}\label{III.1}

Eq.(\ref{cec10}) simplifies the quadratic action for vector perturbations as:
\begin{equation}\label{sc10}
	\begin{aligned}
		S^{(2)}_v= & -\dfrac{1}{2}\sum_A\int d\eta d^3\vec{k} ~a^2\Big[\left(c_2+c_3\right)\beta_A^{\prime2}+2c_2\lambda_A^{\prime2}+2c_4\beta_A'\lambda_A'+4\left(c_2+c_3\right)\mathfrak{p}_A\mathcal{H}k\beta_A\lambda_A\\
		&+\left(c_2+c_3\right)k^2\lambda_A^2+\left(4c_2+2c_3\right)\mathfrak{p}_Ak\beta_A\lambda_A'+c_4\left(-2\mathfrak{p}_Ak\mathcal{H}\beta_A^2-k^2\beta_A\lambda_A+\mathfrak{p}_Ak\mathcal{H}\lambda_A^2\right)\Big].
	\end{aligned}
\end{equation}
The mixing term $2c_4\beta_A'\lambda_A'$ brings difficulty to tell whether the vector perturbations suffer from ghost instability. So the perturbation fields need to be diagonalized. We denote the kinetic matrix, which contains all the coefficients of kinetic terms in (\ref{sc10}), as $\mathbb{M}$ :
\begin{equation}\label{M}
  \mathbb{M}=-\frac{a^2}{2}
  \begin{pmatrix}
		c_2+c_3 & c_4\\
		c_4 & 2c_2
  \end{pmatrix}.
\end{equation}

Therefore, three possible cases of the two eigenvalues about matrix $\mathbb{M}$ are included: 
\begin{itemize}
	\item both two eigenvalues vanish, $i.e.$,  $\det\mathbb{M}=\mathrm{tr}~\mathbb{M}=0$;
	\item one of the eigenvalues vanish, $i.e.$,  $\det\mathbb{M}=0,~\mathrm{tr}~\mathbb{M}\neq0$;
	\item both two eigenvalues are non-zero,  $i.e.$, $\det\mathbb{M}\neq0,~\mathrm{tr}~\mathbb{M}\neq0$.
\end{itemize}
The first case would never bring ghost instability, since no kinetic terms left in the quadratic action after the diagonalization. The second case means that the action after diagonalization will contain up to one kinetic term while the rest perturbation field becomes non-dynamical. In order to exclude ghost modes, we need the coefficient of the kinetic term to be positive after substituting the constraint back into the action. As for the last case, there will be two kinetic terms after the diagonalization and it requires both two eigenvalues are positive to avoid ghost instability, $i.e.$, $\det\mathbb{M}>0,~\mathrm{tr}~\mathbb{M}>0$.

\begin{center}
	$\mathfrak{a.}$ \textbf{The case that both two eigenvalues vanish}
\end{center}

If both two eigenvalues of the kinetic matrix (\ref{M}) vanish, then we have
\begin{equation}\label{vector1a}
		3c_2+c_3=0,\quad
		2c_2\left(c_2+c_3\right)=c_4^2,
\end{equation}
leading to $c_1=c_2=c_3=c_4=0$, which is a trivial case.\\

\begin{center}
	$\mathfrak{b.}$ \textbf{The case that one of the eigenvalues vanishes}  
\end{center}

It requires
\begin{equation}\label{vector1b}
		3c_2+c_3\neq0,\quad
		2c_2\left(c_2+c_3\right)=c_4^2,
\end{equation}
then we will discuss the cases $c_4=0$ and $c_4\neq0$ separately.\\


{\boldmath{$c_4=0$}}: In this case, either $c_2$ or $c_2+c_3$ would vanish. Taking $c_1=0$ and $W>0$ into account, the quadratic action has such from:
\begin{equation}\label{sv1b1}
		S^{(2)}_{v(c_2+c_3=0)}= \sum_A\int d\eta d^3\vec{k}~a^2\left( c_3\mathfrak{p}_Ak\beta_A\lambda_A'+c_3\lambda_A^{\prime2}\right),
\end{equation}
\begin{equation}
	S^{(2)}_{v(c_2=0)}= -\dfrac{1}{2}\sum_A\int d\eta d^3\vec{k}~a^2\left( c_3k^2\lambda_A^2-2c_3\mathfrak{p}_Ak\beta_A'\lambda_A+c_3\beta_A^{\prime2}\right). 
\end{equation}

In the case $c_2+c_3=0$, $\beta_A$ is not dynamical. The variation with respect to $\beta_A$ leads to constraint $\lambda'_A=0$. Substituting it to action (\ref{sv1b1}) result in $S^{(2)}_{v(c_2+c_3=0)}=0$. In the case $c_2=0$ the same program can be done and also lead to the result  $S^{(2)}_{v(c_2=0)}=0$.\\

{\boldmath{$c_4\neq0$}}: To diagnalize the kinetic matrix, we introduce two new fields in terms of $\beta_A$ and $\lambda_A$ base on the eigenvector:
\begin{equation}
		\alpha_{1A}=-2c_2\beta_A+c_4\lambda_A,\quad
		\alpha_{2A}=c_4\beta_A+2c_2\lambda_A,
\end{equation}
and then we can rewrite the action in terms of $\alpha_{1A}$ and $\alpha_{2A}$:
\begin{equation}
	\begin{aligned}
		S^{(2)}_v=&-\dfrac{1}{8c_2^2(3c_2+c_3)^2}\sum_A\int d\eta d^3\vec{k}~a^2 \Big\{2c_2\left(3c_2+c_3\right)\left[\left(c_2+c_3\right)k^2-c_4\mathfrak{p}_A\mathcal{H}k\right]\alpha_{1A}^2 \\
		&+2c_2\left(3c_2+c_3\right)c_4 k^2\alpha_{2A}\alpha_{1A}-4c_2\mathfrak{p}_A\left(2c_2+c_3\right)\left(3c_2+c_3\right)k\alpha_{2A}'\alpha_{1A}+2c2\left(3c_2+c_3\right)^2\alpha_{2A}^{\prime2}\Big\},
	\end{aligned}
\end{equation}
where $\alpha_{1A}$ is non-dynamical, which yells the constraint:
\begin{equation}\label{36}
	4c_2\big[-c_4\mathfrak{p}_A\mathcal{H}+(c_2+c_3)k\big]\alpha_{1A}+2c_2c_4k\alpha_{2A}-4c_2\mathfrak{p}_A(2c_2+c_3)\alpha_{2A}'=0.
\end{equation}

Then $\alpha_{1A}$ can be solved. Since the final result is very complicate, we just write down the kinetic term in the action:
\begin{equation}\label{vac10c4!0}
	S^{(2)}_v=\sum_A\int d\eta d^3\vec{k}~a^2 \left\{\dfrac{\mathfrak{p}_A\left(3c_2+c_3\right)c_4\mathcal{H}+c_2^2k}{2\left(4c_2^2+c_4^2\right)\left[(c_2+c_3)k-\mathfrak{p}_A c_4\mathcal{H}\right]}\alpha_{2A}^{\prime2}+...\right\}.
\end{equation}
In order to exclude ghost modes, condition 
\begin{equation}\label{coe}
	\dfrac{\mathfrak{p}_A\left(3c_2+c_3\right)c_4\mathcal{H}+c_2^2k}{(c_2+c_3)k-\mathfrak{p}_A c_4\mathcal{H}}>0
\end{equation}
must be satisfied when $k \in (0,\infty)$. The condition $2c_1+c_2\geq0$ that keeps the tensor perturbation free of ghost instability, along  with $c_1=0$ leads to $c_2\geq0$. Besides, taking (\ref{vector1b}) into account, we can derive that $c_2>0$ and $c_2+c_3>0$. Therefore, for $k\rightarrow0$, the coefficient above (\ref{coe}) becomes
\begin{equation}
	-\left(3c_2+c_3\right)-\mathfrak{p}_A\frac{\left(2c_2+c_3\right)^2}{c_4\mathcal{H}}k+\mathcal{O}\left(k^2\right).
\end{equation} 
Then we can directly find out that (\ref{coe}) cannot always be positive, which causes ghost instability in (\ref{vac10c4!0}). Thus, in this case the ghost instability cannot be canceled.

\begin{center}
	$\mathfrak{c.}$ \textbf{The case that both the eigenvalues are positive}
\end{center}

In the case that the kinetic matrix is positive definite, we have inequalities 
\begin{equation}\label{38}
		3c_2+c_3<0,\quad
		2c_2\left(c_2+c_3\right)>c_4^2.
\end{equation}

However, taking the condition (\ref{tensor}) into account,  $2c_2\left(c_2+c_3\right)>0$ implies $c_2>0$ and $c_2+c_3>0$, which results in $3c_2+c_3>0$. This condition is in contradiction with (\ref{38}). Therefore, this case also cannot hold.\\

\subsubsection{$c_1\neq0$}\label{III.2}

Since $c_1\neq0$, we can solve for $\gamma_A$ from (\ref{cec10}), and then the quadratic action for vector perturbation can be rewritten as follows:
\begin{equation}\label{39}
	\begin{aligned}
		S^{(2)}_v= &-\sum_A\int d\eta d^3\vec{k}~ \dfrac{a^2}{16c_1} \Big\{\left(8c_1Z+c_4^2\right)\beta_A^{\prime2}+4\left(c_2^2-4c_1^2\right)\lambda_A^{\prime2}+4c_4\left(6c_1-c_2\right)\lambda_A'\beta_A'\\
		&+\left(8c_1Zk+c_4^2k+4c_2c_4\mathfrak{p}_A\mathcal{H}\right)k\lambda_A^2-4k\left[\left(2c_1+c_2\right)c_4k+\left(c_4^2-8c_1Z\right)\mathfrak{p}_A\mathcal{H}\right]\beta_A\lambda_A\\
		&+4k\left[-4c_1^2k-4c_1c_4\mathfrak{p}_A\mathcal{H}+c_2\left(c_2k+c_4\mathfrak{p}_A\mathcal{H}\right)\right]\beta_A^2+2\mathfrak{p}_A\left[8c_1\left(c_2+c_3\right)+4c_2^2-c_4^2\right]k\beta_A\lambda_A'\Big\},
	\end{aligned}
\end{equation}
and the mixing terms $\lambda_A'\beta_A'$ still exist.

\begin{center}
	$\mathfrak{d.}$ \textbf{The case that both two eigenvalues vanish}
\end{center}

After similar operations in $\mathfrak{a}$, we have:
\begin{equation}\label{d}
	\begin{cases}
		4c_2^2+8c_1\left(c_2+c_3\right)+c_4^2=0\\
		2\left(2c_1-c_2\right)\left(2c_1+c_2\right)Z-\left(3c_2-10c_1\right)c_4^2=0
	\end{cases}.
\end{equation}

Taking $W>0$ and (\ref{tensor}) into consider, one can determine that $2c_1+c_2=0$ as well as (\ref{d}) will lead to $c_4=0$ and $c_3=0$, which contradicts $W>0$. Therefore, in this case $2c_1+c_2>0$ always holds, which means the tensor perturbations must be propagating.

\begin{center}
	$\mathfrak{e.}$ \textbf{The case that one of the eigenvalues vanishes}
\end{center}

In such case, we have
\begin{equation}\label{41}
	\begin{cases}
		4c_2^2+8c_1\left(c_2+c_3\right)+c_4^2\neq0\\
		2\left(2c_1-c_2\right)\left(2c_1+c_2\right)Z-\left(3c_2-10c_1\right)c_4^2=0
	\end{cases},
\end{equation}
then we will discuss the cases $\left(3c_2-10c_1\right)c_4^2=0$ and $\left(3c_2-10c_1\right)c_4^2\neq0$ separately.\\

{\boldmath{$c_4=0$}}: The second equation in (\ref{41}) requires $2c_1-c_2=0$ or $2c_1+c_2=0$ or $Z=0$. Moreover, the first inequality prevents any combination of these three cases. After taking each case into account, we can write down the quadratic action separately:
\begin{equation}
	\begin{aligned}
		S^{(2)}_{v(2c_1-c_2=0)}&= -\dfrac{1}{2}\sum_A\int d\eta d^3\vec{k}~a^2\left(4c_1+c_3\right)\left(k^2\lambda_A^2-2\mathfrak{p}_Ak\lambda_A\beta_A'+\beta_A^{\prime2}\right),\\
		S^{(2)}_{v(2c_1+c_2=0)}&= -\dfrac{1}{2}\sum_A\int d\eta d^3\vec{k}~a^2c_3\left(k^2\lambda_A^2-2\mathfrak{p}_Ak\lambda_A\beta_A'+\beta_A^{\prime2}\right),\\
		S^{(2)}_{v(Z=0)}&= -\dfrac{1}{2}\sum_A\int d\eta d^3\vec{k}~\frac{a^2\left(c_2^2-4c_1^2\right)}{2c_1}\left(k^2\beta_A^2+2\mathfrak{p}_Ak\beta_A\lambda_A'+\lambda_A^{\prime2}\right),
		\end{aligned}
\end{equation}
and just like case $\mathfrak{b}$ when $c_4=0$, one can show that the action in all three cases will finally vanish after substituting the constraints back given by the non-dynamical fields. These results are independent of whether $10c_1-3c_2=0$ or not.\\

{\boldmath{$10c_1-3c_2=0,~c_4\neq0$}}: Since $c_1\neq0$, we can  obtain $c_2=10c_1/3,~c_3=-16c_1/3$ from the equation in (\ref{41}). After substituting them into the action (\ref{39}), we have
\begin{equation}
	\begin{aligned}
		S^{(2)}_{v(10c_1-3c_2=0)}=&-\sum_A\int d\eta d^3\vec{k} ~\dfrac{a^2}{144c_1}\Big[9c_4^2\beta_A^{\prime2}+96c_1c_4\beta'\lambda'+256c_1^2\lambda_A^{\prime2}-6c_4\mathfrak{p}_Ak\left(3c_4\beta_A+8c_1\lambda_A\right)\lambda_A'+512c_1^2\mathfrak{p}_Ak\beta_A\lambda_A' \\&+8c_1\left(32c_1k-3c_4\mathfrak{p}_A\mathcal{H}\right)k\beta_A^2-12c_4\left(16c_1k+3c_4\mathfrak{p}_A\mathcal{H}\right)k\beta_A\lambda_A+9c_4\left(c_4k+8c_1\mathfrak{p}_A\mathcal{H}\right)k\lambda_A^2\Big],
	\end{aligned}
\end{equation}
where both $c_1$ and $c_4$ are non-zero. The  determinant of kinetic matrix is zero, indicating that one of the eigenvalues vanishes. Then we introduce two new fields in terms of $\beta_A$ and $\lambda_A$ base on the eigenvector:
\begin{equation}
	\alpha_{1A}=16c_1\beta_A-3c_4\lambda_A	,\quad\alpha_{2A}=3c_4\beta_A+16c_1\lambda_A,
\end{equation}
to diagnalize the kinetic matrix and then the action can be rewritten as follows:
\begin{equation}\label{10c1-3c2=0}
	\begin{aligned}
		S^{(2)}_{v(10c_1-3c_2=0)}=&-\sum_A\int d\eta d^3\vec{k} ~\dfrac{a^2}{144c_1\left(256c_1^2+9c_4^2\right)^2} \Big\lbrace\left[\left(65536c_1^4+9216 c_1^2c_4^2+81c_4^4\right)k+72c_1c_4\left(256c_1^2+27c_4^2\right)\mathfrak{p}_A\mathcal{H}\right]k\alpha_{1A}^2\\&
		-96c_1c_4k\left[\left(256c_1^2-9c_4^2\right)k+144c_1c_4\mathfrak{p}_A\mathcal{H}\right]\alpha_{1A}\alpha_{2A}+2\left(256c_1^2+9c_4^2\right)\left(256c_1^2-9c_4^2\right)\mathfrak{p}_Ak\alpha_{1A}\alpha_{2A}'\\
		&+24c_1c_4\left[-192c_1c_4k+\left(256c_1^2-117c_4^2\right)\mathfrak{p}_A\mathcal{H}\right]k\alpha_{2A}^2+\left(256c_1^2+9c_4^2\right)^2\alpha_{2A}^{\prime2} \Big\rbrace.
	\end{aligned}
\end{equation}

Since the $\alpha_{1A}$ is non-dynamical field, variation with $\alpha_{1A}$ will lead to a constraint. After substituting the constraint back into action (\ref{10c1-3c2=0}), we finally obtain
\begin{small}
	\begin{equation}
		S^{(2)}_{v(10c_1-3c_2=0)}=-\sum_A\int d\eta d^3\vec{k} \frac{ \left[192c_1c_4^2k+\left(256c_1^2+27c_4^2\right)c_4\mathfrak{p}_A\mathcal{H}\right]\alpha_{2A}^{\prime2}-c_4k\left[24c_1c_4\left(7\mathcal{H}^2+4k^2\right)\left(256c_1^2+27c_4^2\right)\mathfrak{p}_A\mathcal{H}k\right]\alpha_{2A}^2}{\left(65536c_1^4+9216c_1^2c_4^2+81c_4^4\right)k+72c_1c_4\left(256c_1^2+27c_4^2\right)\mathfrak{p}_A\mathcal{H}}.
	\end{equation}
\end{small}

After taking $W>0$ into account, we will obtain $c_1>0$. Then if $c_4>0$, the coefficient of $\alpha_{2A}^{\prime2}$ will be negative in right-handed polarization and $c_4<0$ will also lead to the negative coefficient of $\alpha_{2A}^{\prime2}$ in left-handed polarization. Therefore, this case cannot avoid ghost instability and should be excluded.\\

{\boldmath{$10c_1-3c_2\neq0,~c_4\neq0$}}:  Eq.(\ref{41}) implies that
\begin{equation}\label{c4}
	c_4^2=-\dfrac{2\left(2c_1-c_2\right)\left(2c_1+c_2\right)Z}{10c_1-3c_2}.
\end{equation}
To diagnalize the kinetic matrix, we introduce two new fields in terms of $\beta_A$ and $\lambda_A$ base on the eigenvector:
\begin{equation}
		\alpha_{1A}=\left(4c_1^2-c_2^2\right)\beta_A+\frac{c_4}{2}\left( 6c_1-c_2\right) \lambda_A,\quad
		\alpha_{2A}=-\frac{c_4}{2}\left( 6c_1-c_2\right) \beta_A+\left( 4c_1^2-c_2^2\right) \lambda_A,
\end{equation}
and then we can rewrite the action in terms of $\alpha_{1A}$ and $\alpha_{2A}$:
\begin{equation}\label{esv}
	S^{(2)}_v=\sum_A\int d\eta d^3\vec{k}~\frac{a^2}{V_{00}}\left(V_{11} \alpha_{1A}^2+V_{12}\alpha_{1A}\alpha_{2A}+V_{21}\alpha_{1A}\alpha_{2A}'+V_{22}\alpha_{2A}^2+\tilde{V}_{22}\alpha_{2A}^{\prime2}\right),
\end{equation}
where the coefficients are
\begin{equation}
	\begin{aligned}
		V_{00}=&-4c_1\left(4c_1^2-c_2^2\right)\left[8c_1^3-36c_1^2c_2-10c_1c_2^2 +5c_2^3-\left(6c_1-c_2\right)^2c_3\right]^2,\\ V_{11}=&-k\Big\{4c_1\left(10c_1-3c_2\right)\big[344c_1^3-44c_1^2\left(c_2-3c_3\right)+3c_2^2\left(5c_2+3c_3\right)-2c_1c_2\left(39c_2\right.\\
		&\left.+38c_3\right)\big]c_4\mathfrak{p}_A\mathcal{H}+\left[\left(2c_1+c_2\right)^2\left(1936c_1^4-2976c_1^3c_2+1912c_1^2c_2^2-520c_1c_2^3+49c_2^4\right)\right.\\
		&\left.+2\left(6c_1-c_2\right)\left(2c_1+c_2\right)\left(136c_1^3-84c_1^2c_2+38c_1c_2^2-7c_2^3\right)c_3+\left(6c_1-c_2\right)^4c_3^2\right]k\Big\},\\ V_{12}=&4\left(10c_1-3c_2\right)\left(2c_1-c_2\right)\Big\{8c_1\mathfrak{p}_A\left(14c_1-5c_2\right)\left(2c_1+c_2\right)\left(2c_1+c_2+c_3\right)\mathcal{H}+\left[152c_1^3\right.\\
		&\left.-12c_1^2c_2-30c_1c_2^2+7c_2^3+(6c_1-c_2)^2c_3\right]c_4k\Big\},\\
		V_{21}=&2\left[8c_1^3-36c_1^2c_2-10c_1c_2^2+5c_2^3-\left(6c_1-c_2\right)^2c_3\right]\Big[8c_1^3+6c_1c_2\left(c_2-2c_3\right)\\
		&+44c_1^2\left(c_2+c_3\right)-c_2^2\left(7c_2+c_3\right)\Big]\mathfrak{p}_Ak,\\
		V_{22}=&4k\left(10c_1-3c_2\right)\Big\{\left[-360c_1^4+4c_1^3\left(31c_2-15c_3\right)+c_2^3\left(5c_2-c_3\right)+5c_1c_2^2\left(-7c_2+c_3\right)\right.\\
		&\left.+2c_1^2c_2\left(31c_2+8c_3\right)\right]c_4\mathfrak{p}_A\mathcal{H}+2\left(6c_1-c_2\right)\left(-2c_1+c_2\right)^2\left(2c_1+c_2\right)\left(2c_1+c_2+c_3\right)k\Big\},\\
		\tilde{V}_{22}=&-\Big[8c_1^3-2c_1c_2\left(5c_2-6c_3\right)+c_2^2\left(5c_2-c_3\right)-36c_1^2\left(c_2+c_3\right)\Big]^2, 
	\end{aligned}
\end{equation}
After substituting $\alpha_{1A}$ and $\alpha_{2A}$ into the quadratic action for vector perturbations, we find $\alpha_{1A}$ is non-dynamical, and variation with $\alpha_{1A}$ yields:
\begin{equation}
	2\alpha_{1A}V_{11}+V_{12}\alpha_{2A}+V_{21}\alpha_{2A}'=0.
\end{equation}

One can solve for $\alpha_{1A}$ and substitute it back into the action (\ref{esv}), and it  ultimately yields:
\begin{equation}\label{47}
	S^{(2)}_v= \sum_A\int d\eta d^3\vec{k}~a^2\left[ \dfrac{d_1\mathcal{H}-\mathfrak{p}_Ad_2k}{2\left( d_3d_1\mathcal{H}+\mathfrak{p}_A d_4k\right) }\alpha_{2A}^{\prime2}+\dots \right] .
\end{equation}

Again, we just write down the kinetic term in the final action since the full expression is complicate and we've defined:
\begin{equation}
		\begin{aligned}
		d_1=&-2\left(10c_1-3c_2\right)\left[\left(2c_1+c_2\right)\left(172c_1^2-108c_1c_2+15c_2^2\right)+\left(22c_1-9c_2\right)\left(6c_1-c_2\right)c_3\right]c_4,\\
		d_2=&-8\left(4c_1^2-c_2^2\right)\left(6c_1-5c_2-c_3\right)\left(2c_1+c_2+c_3\right)\left(10c_1-3c_2\right),\\
		d_3=&2c_1\left(4c_1^2-c_2^2\right),\\
		d_4=&\left(4c_1^2-c_2^2\right)\left[\left(2c_1+c_2\right)^2\left(1936c_1^4-2976c_1^3c_2+1912c_1^2c_2^2-520c_1c_2^3+49c_2^4\right)\right.\\
		&\left.+2\left(6c_1-c_2\right)\left(2c_1+c_2\right)\left(136c_1^3-84c_1^2c_2+38c_1c_2^2-7c_2^3\right)c_3+\left(-6c_1+c_2\right)^4c_3^2\right],
	\end{aligned}
\end{equation}
for simplification. Whether the quadratic action contains ghost modes depends on the values of $d_I~(I=1,2,3,4)$. Therefore, we must analysis all possible cases separately.\\

If $d_{I=1,2,3,4}\neq0$: One can tell that $d_3\neq0$ since none of $c_1$ and $2c_1\pm c_2$ vanishes. In order to avoid ghost instability in this case, we need:
\begin{equation}
		d_{I=1,2,3,4}\neq0,\quad
		d_3>0,\quad
		d_4+d_3d_2=0.
\end{equation}
The requirement $d_3>0$ will lead to two different case: 
\begin{equation}
	c_1>0,~-2c_1<c_2<2c_1\quad\mathrm{or}\quad c_1<0,~c_2>-2c_1.
\end{equation}

One can justify that the result for $c_1>0,~-2c_1<c_2<2c_1$ is negative definite. As for the case $c_1<0$, as we will see in next subsection, will lead to ghost instability in scalar modes. Therefore, this case should be excluded.\\

If $d_1=0,\ d_2=0,\ d_4\neq0$: The quadratic action for vector perturbations are 
\begin{equation}
	\mathfrak{p}_A=\pm1:\ S^{(2)}_{v(d1=d2=0)}=0.
\end{equation}

We can solve for $c_3$ from $d_2=0$ to be $c_3=6c_1-5c_2$. Substituting this expression of $c_3$ into $d_1=0$, we will obtain the equation $284c_1^2-148c_1c_2+15c_2^2=0$. To solve this equation, we have
\begin{equation}
	c_2=\frac{2}{15}\left(37\pm4\sqrt{19}\right)c_1.
\end{equation}
From the requirements of background $W>0$, tensor (\ref{tensor}) and vector sector, the allowed region for $c_1$ and $c_2$ is given by these inequalities:

\begin{equation}\label{fg1}
	\begin{cases}
		2c_1+c_2>0\\
		10c_1-7c_2<0\\
		10c_1-3c_2<0\\
		2c_1-c_2\neq0\\
		c_1\neq0
	\end{cases},
\end{equation}
as well as $d_4\neq0$. We can see that the solution $c_2=\frac{2}{15}\left(37+4\sqrt{19}\right)c_1$ happens to locate in the allowed region (\ref{fg1}) when $c_1>0$. Therefore, in this case we have 
\begin{equation}
	c_1>0,~c_2=\frac{2}{15}\left(37+4\sqrt{19}\right)c_1,~c_3=6c_1-5c_2,
\end{equation}
and $c_4$ is restrained by eq.(\ref{c4}).\\

\begin{figure}[htbp]
	\centering
	\includegraphics[scale=0.7]{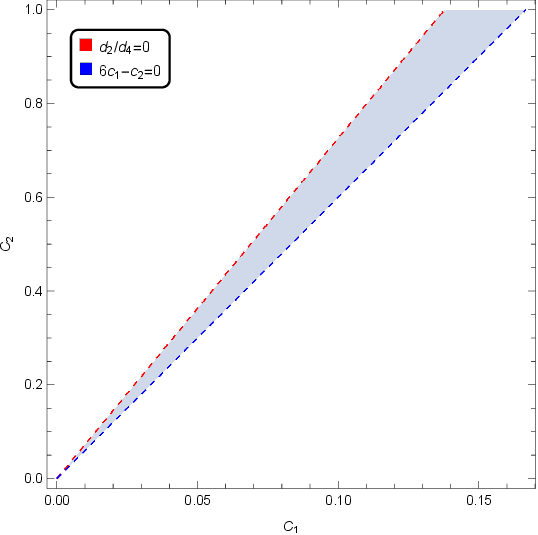}
	\caption{The allowed region of the case $d_1=0,\ d_2/d_4<0$.}
	\label{e}
\end{figure}

If $d_1=0,\ d_2/d_4<0$: In this case, the quadratic action (\ref{47}) has the following form:
\begin{equation}
	\mathfrak{p}_A=\pm1:\ S^{(2)}_{v(d1=0)}= \sum_A\int d\eta d^3\vec{k}~a^2\left[\ -\dfrac{d_2}{2d_4}\alpha_{2A}^{\prime2}+\dots~\right].
\end{equation}
If $\left(22c_1-9c_2\right)\left(6c_1-c_2\right)$ in $d_1$ equal to zero, one can substitute the solution $c_2=22c_1/9$ or $c_2=6c_1$ back into $d_1$, which result in  $d_1\propto c_1^3\neq0$, and it contradicts $d_1=0$. So we can obtain the  $c_3$ from $d_1=0$ in terms of $c_1$ and $c_2$:
\begin{equation}
	c_3=-\dfrac{\left(2c_1+c_2\right)\left(172c_1^2-108c_1c_2+15c_2^2\right)}{\left(22c_1-9c_2\right)\left(6c_1-c_2\right)}.
\end{equation}
 Substituting $c_3$ into the equation of (\ref{41}), we also obtain $c_4^2$, expressed in terms of $c_1$ and $c_2$:
\begin{equation}
	c_4^2=\dfrac{4\left(4c_1^2-c_2^2\right)^2}{\left(22c_1-9c_2\right)\left(6c_1-c_2\right)}.
\end{equation}
Now we have the allowed region:

\begin{equation}
	\begin{cases}
		\left(284c_1^2-148c_1c_2+15c2^2\right)\left(38c_1^2-25c_1c_2+4c_2^2\right)<0\\
		\left(22c_1-9c_2\right)\left(6c_1-c_2\right)>0\\
		96c_1^2-62c_1c_2+9c_2^2<0\\
		c_1\neq0\\
		10c_1-3c_2\neq0\\
		2c_1-c_2\neq0
	\end{cases},
\end{equation}
and we plot the region on Fig.\ref{e}. Within the region, the number of vector perturbations DoFs  is two, and these DoFs are propagating and free of ghost instability.

If $d_2=0,\ d_4=0,\ d_3>0$: The quadratic action in this case is
\begin{equation}
	\mathfrak{p}_A=\pm1:\ S^{(2)}_{v(d2=d4=0)}= \sum_A\int d\eta d^3\vec{k}~a^2\left[\dfrac{1}{2d_3}\alpha_{2A}^{\prime2}+\dots~\right],
\end{equation}
where $d_3=2c_1\left(4c_1^2-c_2^2\right)$ should be positive to avoid ghosts. On can easily derive $c_3=6c_1-5c_2$ from $d_2=0$ and then substitute $c_3$ into (\ref{41}) to obtain $c_4^2$:
\begin{equation}
	c_4^2=-\frac{8\left(2c_1-c_2\right)^2\left(2c_1+c_2\right)}{10c_1-3c_2}.
\end{equation}
The allowed region can be simplified as
\begin{equation}
	\begin{cases}
		c_1<0\\
		2c_1+c_2>0
	\end{cases};
\end{equation}
while substituting $c_3=6c_1-5c_2$ into $d_4=0$ we will obtain $136\,c_1^3-124\,c_1^2c_2+30\,c_1c_2^2-c_2^3=0$. To solve this equation, we have
\begin{equation}\label{d4=0}
	c_1=\frac{c_2}{2},~\text{or}~c_1=\frac{7\pm4\sqrt{2}}{34}c_2.
\end{equation}
In the allowed region, $c_2>0$ always holds, so the solutions (\ref{d4=0}) always satisfy $c_1>0$, which do not locate in the allowed region. So this case should be excluded.

\begin{center}
	$\mathfrak{f.}$ \textbf{The case that both the eigenvalues are positive}
\end{center}

\begin{figure}[htbp]
	\centering
	\subfigure[$~c_1>0$]{\includegraphics[scale=0.437]{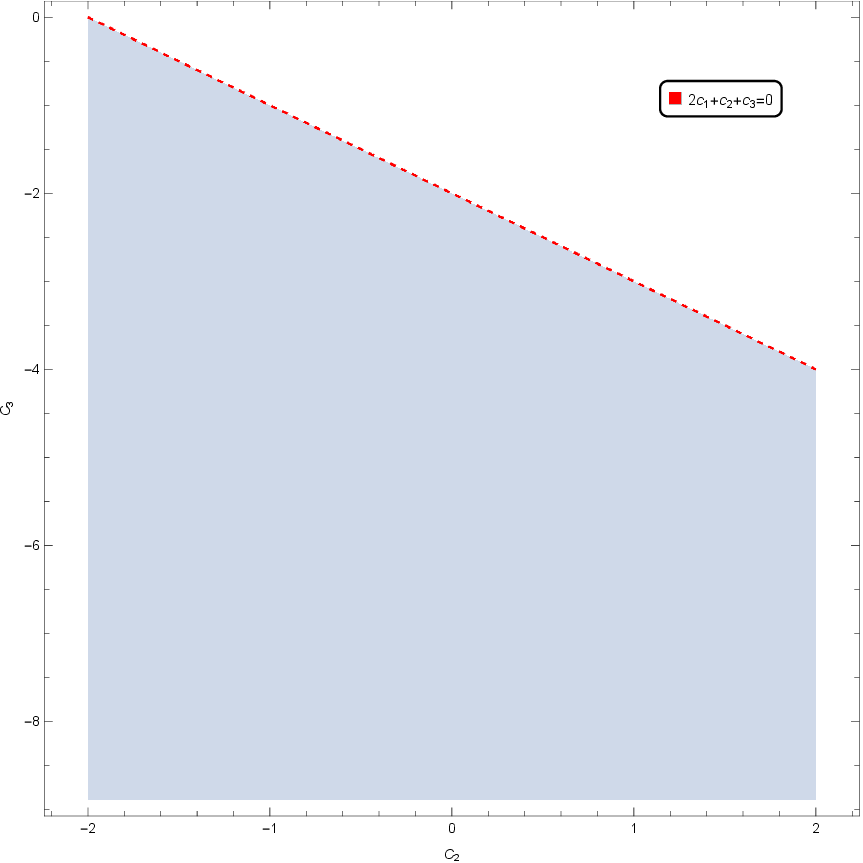}}$\qquad$
	\subfigure[$~c_1<0$]{\includegraphics[scale=0.44]{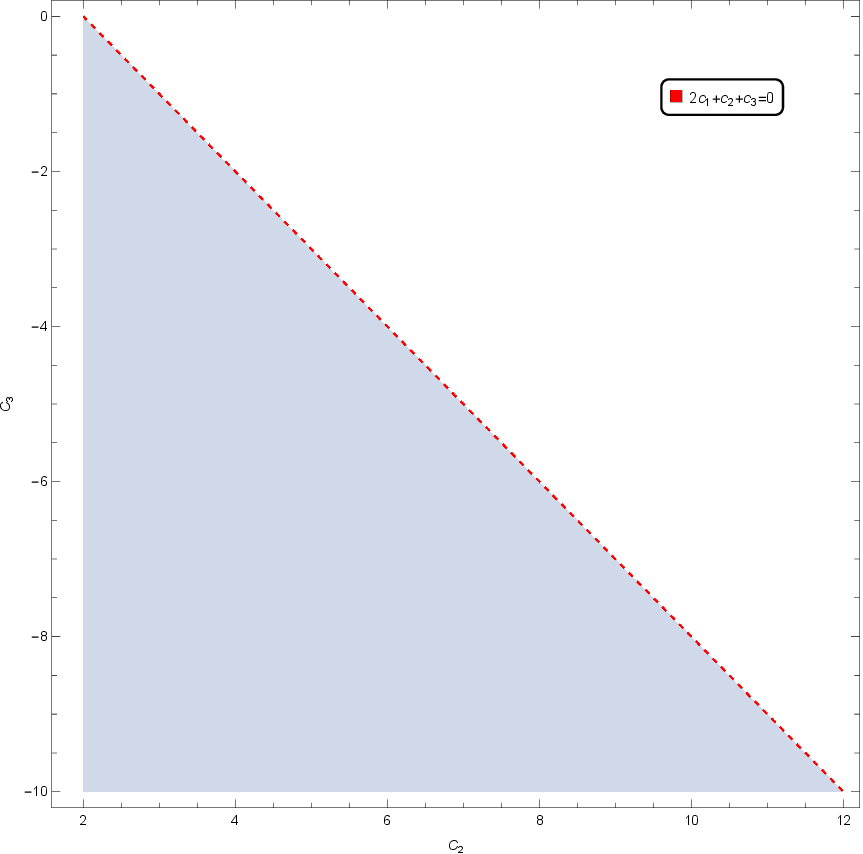}}
	\caption{The allowed region of case where both the eigenvalues are positive, $c_2$ and $c_3$ are normalized with $c_1=\pm1$.}
	\label{f}
\end{figure}

The requirement of this case yields:
\begin{equation}\label{56}
	\frac{4c_2^2+8c_1\left(c_2+c_3\right)+c_4^2}{8c_1}<0,\quad
	\frac{2\left(2c_1-c_2\right)\left(2c_1+c_2\right)Z-\left(3c_2-10c_1\right)c_4^2}{2c_1}<0.
\end{equation}
Normalize $c_2$, $c_3$ and $c_4$ with $c_1$, we can plot the allow region for $c_2$ and $c_3$, which is in agreement with $W>0$, (\ref{tensor}) and (\ref{56}), in Fig.\ref{f}.

\subsection{Quadratic Action for Scalar Perturbations}

The quadratic action for scalar perturbations in Fourier space is:
\begin{equation}\label{sa1}
	\begin{aligned}
		S^{(2)}_s=&-\int d\eta d^3\vec{k}~a^2 \left\{-2k^2\mathcal{H}WAB+\left(2\mathcal{H}^2+\mathcal{H}'\right)WA^2-\frac{k^2Z}{2}\left(k^2B^2+2k^2B\beta+k^2\beta^2\right)+\left(c_2-2c_1\right)k^2\lambda^{\prime2}\right.\\
		&\left.+\frac{k^2Z}{2}\left(A^2-2A\beta'+\beta^{\prime2}\right)+\left(2c_1-c_2\right)k^4\lambda^2+c_4k^2\left[\beta'\lambda'-k^2\left(B+\beta\right)\lambda-A\lambda'\right] \right\},
	\end{aligned}
\end{equation}
and we also marked $A^*B=AB$, $A^*A=A^2$, and so on, for simplicity. No matter what values $c_i,\left(i=1,2,3,4\right) $ takes, $A$ and $B$ are non-dynamical fields. Variation with these fields yields two constraints:
\begin{equation}\label{sc1}
	\begin{aligned}
		2k^2W\mathcal{H}B-2\left(2\mathcal{H}^2+\mathcal{H}'\right)WA-Zk^2\left(A-\beta'\right)+c_4k^2\lambda'=0,\\
		2W\mathcal{H}A+Zk^2\left(B+\beta\right)+c_4k^2\lambda=0.
	\end{aligned}
\end{equation}

Then we can solve for $A$ and $B$ from equations (\ref{sc1}), and substitute them into action (\ref{sa1}). After that, we have
\begin{equation}\label{sa2}
	\begin{aligned}
		S^{(2)}_s=&-\int d\eta d^3\vec{k}~ \frac{a^2}{2X} \Big\{2k^2WZ\left[2\mathcal{H}^2\left(W+Z\right)+Z\mathcal{H}'\right]\beta^{\prime2}+\left[2k^2X\left(c_2-2c_1\right)-c_4^2k^4Z\right]\lambda^{\prime2}\\
		&+4c_4k^2W\left[2\mathcal{H}^2\left(W+Z\right)+Z\mathcal{H}'\right]\beta'\lambda'+4k^4\mathcal{H}WZ^2\,\beta\beta'+4c_4\mathcal{H}k^4WZ\,\lambda\beta'+4k^4c_4^2\mathcal{H}W\,\lambda\lambda'\\
		&+4c_4\mathcal{H}k^4WZ\,\beta\lambda'+k^4\left[c_4^2\left(2W\mathcal{H}'+4\mathcal{H}^2W+k^2Z\right)+2\left(2c_1-c_2\right)X\right]\lambda^2-4\mathcal{H}^2k^4W^2Z\,\beta^2\\
		&-8c_4k^4\mathcal{H}^2W^2\,\beta\lambda \Big\},
	\end{aligned}
\end{equation}
where $X=k^2Z^2+4\mathcal{H}^2W\left(W+Z\right)+2WZ\mathcal{H}'$, and mixing term $\beta'\lambda'$ also exist.\\

\subsubsection{The case that both two eigenvalues vanish}

In this case, one can have $\mathrm{tr}~\mathbb{M}=\det\mathbb{M}=0$, where $\mathbb{M}$ is the kinetic matrix of the quadratic action (\ref{sa2}). The requirement $\det\mathbb{M}=0$  corresponding to:
\begin{equation}\label{sdetm1}
	\dfrac{a^4k^4W\left(c_4^2+4c_1Z-2c_2Z\right)\left[2\mathcal{H}^2\left(W+Z\right)+Z\mathcal{H}'\right]}{2X}=0.
\end{equation}

Remember that $W>0$ and $2\mathcal{H}^2\left(W+Z\right)+Z\mathcal{H}'$ cannot always vanish for $\mathcal{H}$ is changing with time. Therefore, if the equation (\ref{sdetm1}) always holds, the following equation will be necessary:
\begin{equation}\label{sr1}
	c_4^2+2Z\left(2c_1-c_2\right)=0.
\end{equation}

Then the equation $\mathrm{tr}~\mathbb{M}=0$ can be simplified as follows:
\begin{equation}\label{strm1}
	\dfrac{a^2k^2W\left(4c_1-2c_2-Z\right)\left[2\mathcal{H}^2\left(W+Z\right)+Z\mathcal{H}'\right]}{X}=0,
\end{equation}
and this equation leads to
\begin{equation}\label{sr2}
	4c_1-2c_2-Z=0.
\end{equation}

From the equation (\ref{sr1}) and (\ref{sr2}), along with the requirements given by background $W>0$ and tensor sector (\ref{tensor}), we can derive $c_4=0,~c_2=2c_1,~c_3=-4c_1~\mathrm{and}~c_1>0$. This is just the case of TEGR and is compatible with the case $\mathfrak{d}$ in vector sector, where quadratic action for vector perturbations vanishes. Therefore, the DoFs of vector and scalar perturbations in the tetrad are not propagating, leaving only two tensor perturbation DoFs that are still dynamical and free from ghost instability.\\

\subsubsection{The case that one of the eigenvalues vanishes}

In this case, we have $\det\mathbb{M}=0$ and $\mathrm{tr}~\mathbb{M}\neq0$. From (\ref{sdetm1}) and (\ref{strm1}) we have:
\begin{equation}\label{65}
	c_4^2+2Z\left(2c_1-c_2\right)=0,\quad
	4c_1-2c_2-Z\neq0,
\end{equation}
and we are going to discuss $c_4=0$ and $c_4\neq0$ separately. Besides, $c_4=0$ will lead to $Z=0$ or $2c_1-c_2=0$ and the inequality in (\ref{65}) prevents the combination of them.\\

\begin{figure}[htbp]
	\centering
	\subfigure[$~c_4=0,\ c_3=-2c_1-c_2,\ -2c_1<c_2<2c_1$]{\label{figscalar1}\includegraphics[scale=0.44]{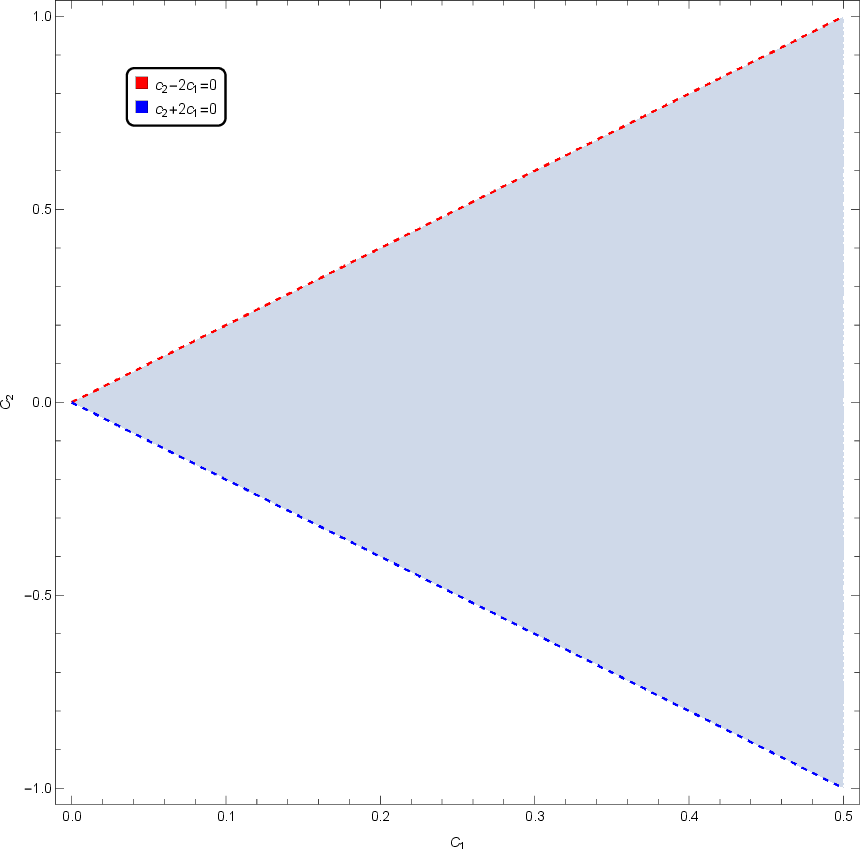}}$\qquad$
	\subfigure[$~c_4=0,\ c_2=2c_1,\ c_1\geq0,\ c_3<-4c_1$]{\label{figscalar2}\includegraphics[scale=0.44]{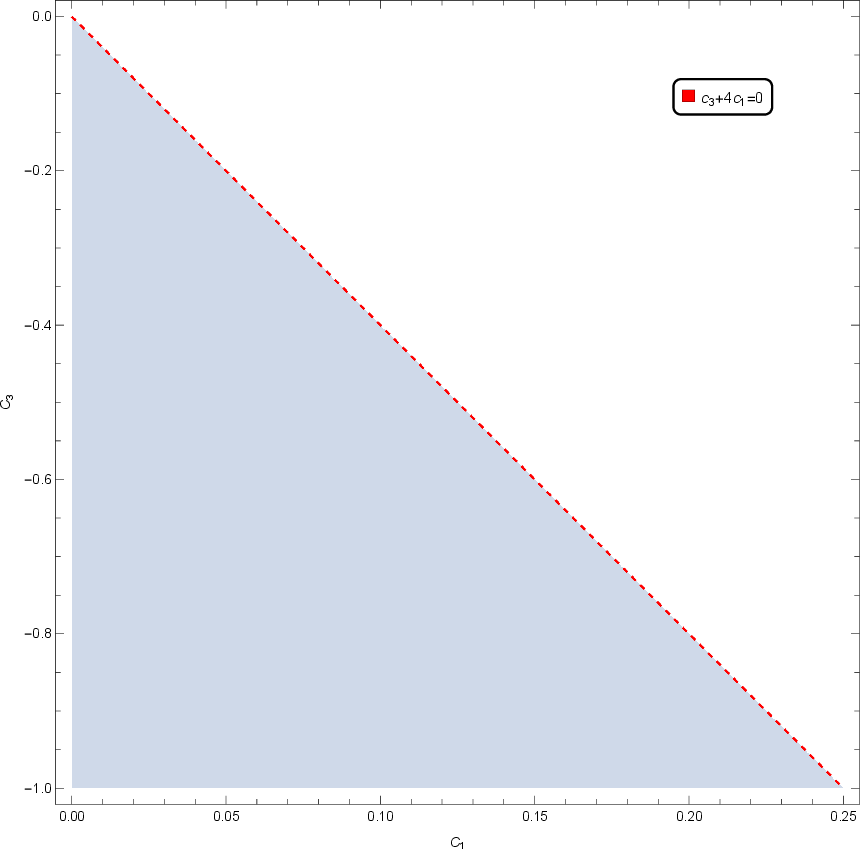}}
	\caption{(a): The allowed region of case $c_4=0,~Z=0,~2c_1-c_2\neq0$. (b): The allowed region of case $c_4=0,~2c_1-c_2=0,~Z\neq0$}
\end{figure}

{\boldmath{$\mathrm{If}~c_4=0,~Z=0,~2c_1-c_2\neq0$}}: After combining the requirements from background $W>0$ and tensor sector (\ref{tensor}), we have $c_3=-2c_1-c_2$ and $2c_1+c_2>0$. Then the quadratic action for scalar perturbations (\ref{sa2}) can be simplified to
\begin{equation}
	S^{(2)}_s=\int d\eta d^3\vec{k}~a^2k^2 \left(2c_1-c_2\right)\left[\lambda^{\prime2}-k^2\lambda^2\right],
\end{equation}
Therefore, the ghost modes in scalar perturbations are canceled iff $2c_1-c_2>0$. This case is compatible with the case $S^{(2)}_{v(Z=0)}$ in $\mathfrak{e}$ in vector sector, where the DoFs of vector perturbations are not dynamical. Now we have the allowed region:
\begin{equation}
	\begin{cases}
		-2c_1<c_2<2c_1\\
		c_3=-2c_1-c_2\\
		c_4=0\\
		c_1>0
	\end{cases},
\end{equation}
and we plot the region on Fig.\ref{figscalar1}. Within the region, the number of tensor and scalar perturbations DoFs are two and one.\\

{\boldmath{$\mathrm{If}~c_4=0,~2c_1-c_2=0,~Z\neq0$}}: The requirements given by $W>0$ and $2c_1+c_2\geq0$, along with $2c_1-c_2=0$ and $Z\neq0$ will result in
\begin{equation}\label{sc402}
	c_2=2c_1,\quad4c_1+3c_3<0,\quad4c_1+c_3\neq0,\quad c_1\geq0.
\end{equation}
Then the quadratic action for scalar perturbations (\ref{sa2}) can be rewritten as follows:
\begin{equation}
	\begin{aligned}
		S^{(2)}_s=\int d\eta d^3\vec{k}~ &\frac{a^2k^2\left(4c_1+c_3\right)\left(4c_1+3c_3\right)}{2X} \Big\{ \left[\left(4c_1-c_3\right)\mathcal{H}^2+\left(4c_1+c_3\right)\mathcal{H}'\right]\beta^{\prime2}+2\left(4c_1+c_3\right)\mathcal{H}k^2\beta\beta'\\
		&+\left(4c_1+3c_3\right)\mathcal{H}^2k^2\beta^2 \Big\}.
	\end{aligned}
\end{equation}	

In order to exclude ghost modes, condition
\begin{equation}\label{sc3}
	O_1=\frac{a^2k^2\left(4c_1+c_3\right)\left(4c_1+3c_3\right)\left[\left(4c_1-c_3\right)\mathcal{H}^2+\left(4c_1+c_3\right)\mathcal{H}'\right]}{2\left[-\left(4c_1-c_3\right)\left(4c_1+3c_3\right)\mathcal{H}^2+\left(4c_1+c_3\right)^2k^2-\left(4c_1+c_3\right)\left(4c_1+3c_3\right)\mathcal{H}'\right]}>0
\end{equation}
must be satisfied when $k\in(0,\infty)$. When $k\rightarrow0$, we have $O_1\simeq-a^2(4c_1+c_3)k^2/2$. So if ghost instability is avoided when $k\rightarrow0$, condition $4c_1+c_3<0$ is necessary. On the other hand, when $k\rightarrow+\infty$, the coefficient 
\begin{equation}
	O_1\simeq\frac{a^2\left(4c_1+3c_3\right)\left[\left(4c_1-c_3\right)\mathcal{H}^2+\left(4c_1+c_3\right)\mathcal{H}'\right]}{2\left(4c_1+c_3\right)},
\end{equation}
so in order to avoid ghost instability, $\left(4c_1-c_3\right)\mathcal{H}^2+\left(4c_1+c_3\right)\mathcal{H}'>0$ is also necessary. One can also examine that when $4c_1+c_3<0$ and $\left(4c_1-c_3\right)\mathcal{H}^2+\left(4c_1+c_3\right)\mathcal{H}'>0$ are satisfied at the same time, the inequality $O_1>0$ (\ref{sc3}) indeed holds. Besides, the condition $c_1\geq0$ along with $c_2=2c_1$ ensure that tensor perturbation is free of ghost instability. Such case is compatible with the cases $S^{(2)}_{v(c_2=0)}$ with $c_1=0$ and $S^{(2)}_{v(2c_1-c_2=0)}$ with $c_1>0$ in vector sector, leading to non-dynamical vector perturbations. If $c_1=0$, the tensor perturbations are not propagating and the number of DoFs of scalar perturbations is one. On the other hand, when $c_1>0$, the tensor perturbations are both dynamical while the DoFs in scalar part remains the same. Finally, we can put these cases together and draw the allowed region 
\begin{equation}
	\begin{cases}
		c_4=0\\
		c_2=2c_1\\
		4c_1+c_3<0\\
		c_1\geq0
	\end{cases}
\end{equation}
for $c_1$ and $c_2$ on Fig.\ref{figscalar2}.\\

{\boldmath{$\mathrm{If}~c_4^2=2Z\left(c_2-2c_1\right)>0$}}: In order to diagnalize the kinetic matrix, two new fields in terms of $\beta$ and $\lambda$ will be introduced:
\begin{equation}
		\alpha_1=c_4\beta-Z\lambda,\quad
		\alpha_2=Z\beta+c_4\lambda.
\end{equation}

Then the quadratic action for scalar perturbation (\ref{sa2}) can be rewritten in terms of $\alpha_1$ and $\alpha_2$ with only one kinetic term:
\begin{equation}
		S^{(2)}_s=-\int d\eta d^3\vec{k}~\frac{a^2\left(4c_1-2c_2-Z\right)^2}{\left(c_4^2+Z^2\right)^2} \frac{ZWk^2}{X}\Big\{\left[2\mathcal{H}^2\left(W+Z\right)+Z\mathcal{H}'\right]\alpha_2^{\prime2}+2k^2\mathcal{H}Z\alpha_2\alpha_2'-2W\mathcal{H}^2k^2\alpha_2^2  \Big\}.
\end{equation}
To avoid ghost instability, we need 
\begin{equation}
	O_2=\frac{ZWk^2\left[2\mathcal{H}^2\left(W+Z\right)+Z\mathcal{H}'\right]}{k^2Z^2+4\mathcal{H}^2W\left(W+Z\right)+2WZ\mathcal{H}'}<0
\end{equation}
to be satisfied when $k\in(0,\infty)$.When $k\rightarrow0$, we have $O_2\simeq Zk^2/2$. So if ghost instability is avoided when $k\rightarrow0$, condition $Z<0$ is necessary. On the other hand, when $k\rightarrow+\infty$, the coefficient 
\begin{equation}\label{sc4}
	O_2\simeq\frac{W\left[2\mathcal{H}^2\left(W+Z\right)+Z\mathcal{H}'\right]}{Z},
\end{equation}
so in order to avoid ghost instability, $2\mathcal{H}^2\left(W+Z\right)+Z\mathcal{H}'>0$ is also necessary. One can also check that when $Z<0$ and $2\mathcal{H}^2\left(W+Z\right)+Z\mathcal{H}'>0$ are satisfied at the same time, the inequality $O_2<0$ (\ref{sc4}) indeed holds. However, when taking the vector sector into account, there is no compatible case that is in agreement with the requirements given above. Therefore, in this case, the ghost instability in vector, scalar and tensor perturbations cannot be avoided at the same time.\\

\subsubsection{The case that both the eigenvalues are positive}

In order to avoid ghost instability in this case, we need $\mathrm{tr}~\mathbb{M}>0$ and $\det\mathbb{M}>0$ when $k\in(0,\infty)$. Therefore we have:
\begin{align}
		-\frac{a^4k^4W\left(c_4^2+4c_1Z-2c_2Z\right)\left[2\mathcal{H}^2\left(W+Z\right)+Z\mathcal{H}'\right]}{2k^2Z^2+8\mathcal{H}^2W\left(W+Z\right)+4WZ\mathcal{H}'}>0,\label{sdetm2}\\
		\frac{a^2k^4Z\left(c_4^2+4c_1Z-2c_2Z\right)-a^2k^2\left(Z-4c_1+2c_2\right)\left[4W(W+Z)\mathcal{H}^2+2WZ\mathcal{H}'\right]}{2k^2Z^2+8\mathcal{H}^2W\left(W+Z\right)+4WZ\mathcal{H}'}>0,\label{strm2}
\end{align}
along with $W>0$ given by background equation and (\ref{tensor}) from tensor sector. Similar to the cases in the former subsubsections, these conditions result in
\begin{equation}\label{sr3}
	-2c_1\leq c_2<2c_1,\quad Z<0,\quad c_4^2<-Z(2c_1-c_2).
\end{equation} 

If $c_2=-2c_1$, (\ref{sr3}) leads to $c_3<0$, $c_1>0$ and $c_4^2<-8c_1c_3$, where tensor perturbations are not propagating and is compatible with the case $S^{(2)}_{v(2c_1+c_2=0)}$ in vector sector, which are also not dynamical, leaving two propagating DoFs in scalar perturbations and free of ghost instability. Besides, one should remember that the case $S^{(2)}_{v(2c_1+c_2=0)}$ in vector sector requires $c_4=0$.

As for $-2c_1<c_2<2c_1$ and $c_1>0$, the number of tensor perturbations DoFs is two and is compatible with case $\mathfrak{f}$ in vector sector. After combining the requirements given by vector sector, one can finally determine the range of values allowed for $c_i\left(i=1,2,3,4\right) $:
\begin{equation}
	-2c_1<c_2<2c_1,\quad Z<0,\quad-\left|m\right|<c_4<\left|m\right|,
\end{equation} 
where $m^2= Min\left\{2Z\left(-2c_1+c_2\right),~-4c_2^2-8c_1\left(c_2+c_3\right),~-2\left(4c_1^2-c_2^2\right)Z/\left(10c_1-3c_2\right)\right\}$. In this case, all of the two DoFs of tensor perturbations, four DoFs of vector perturbations and two DoFs of scalar perturbations are dynamical and free of ghost instability.\\

\section{The Degenerate Case : \boldmath{$W=0$}}\label{IV}

In this section, we will analysis the perturbations when $W=0$. According to the background equation (\ref{frd}), both $\rho$ and $p$ vanish once $W=0$. Although the matter is absent now, the Hubble parameter $\mathcal{H}$ can still evolve over time, which is a nontrivial vacuum solution of the extended NGR model.

\subsection{Quadratic Action for Tensor Perturbations}
The quadratic action for tensor perturbations with $W=0$ exhibits a congruent structure with that of  (\ref{26}) since it does not explicitly depend on $W$:
\begin{equation}
	\begin{aligned}
		S^{(2)}_{T(W=0)}&=\int d\eta d^3\vec{k}~\frac{a^2}{8}  \left[-\left(2c_1+c_2\right)\eta^{\alpha\beta}\partial_\alpha h^{T}_{ij}\partial_\beta h^{T}_{ij}+3c_4\mathcal{H}\epsilon^{i j k}h^{T}_{il}\partial_jh^{T}_{kl}\right]\\&=\sum_A \int d\eta d^3\vec{k}~\frac{a^2}{4}\left[ \left(2c_1+c_2\right)h^{\prime2}_A-k^2\left( 2c_1+c_2-\frac{3c_4\mathcal{H}\mathfrak{p}_A}{k}\right) h^2_A\right].
	\end{aligned}
\end{equation}
It is clear that the condition $2c_1+c_2\geq0$ still arises to avoid the ghost instability in tensor perturbations.

\subsection{Quadratic Action for Vector Perturbations}
The quadratic action for vector perturbations in Fourier space is equivalent to the action given in (\ref{28}) since it is also independent on $W$:
\begin{equation}
	\begin{aligned}
		S^{(2)}_{v(W=0)}=&-\dfrac{1}{2}\sum_A\int d\eta d^3\vec{k}~ a^2\Big[Z\beta^{\prime2}_A+k^2Z\lambda_A^2+4\mathfrak{p}_AkZ\mathcal{H}\beta_A\lambda_A-2c_1k^2\beta_A^2+2c_2k^2\beta_A\gamma_A\\
		&-2c_1k^2\gamma_A^2+\left(2c_2-4c_1\right)\lambda^{\prime2}_A+\mathfrak{p}_Ak\left(2c_2-4c_1\right)\gamma_A\lambda_A'+\left(4c_2+2c_3\right)\mathfrak{p}_Ak\beta_A\lambda_A'\\
		&+c_4\left(2\beta_A'\lambda_A'-k^2\beta_A\lambda_A-k^2\lambda_A\gamma_A-2\mathfrak{p}_Ak\mathcal{H}\beta_A^2+\mathfrak{p}_Ak\mathcal{H}\lambda_A^2-\mathfrak{p}_Ak\gamma_A\beta_A'\right)\Big].
	\end{aligned}
\end{equation}

The analysis is the same as what we have presented in Section \ref{III}, with the additional restriction that $W=0$. The cases $c_2+c_3=0$ and $c_2=0$ in $\mathfrak{b}$ along with $W=0$ will result in $c_1=c_2=c_3=c_4=0$, while the other cases under the condition $c_1=0$ have been excluded in Subsubsection \ref{III.1}. Besides, the case $c_1\neq0$ still needs discussion. Through a series of calculations, it can be determined that there are two cases for vector perturbations that are free from ghosts:
\begin{equation}\label{vw0}
	c_4=0,~c_2=2c_1,~c_3=-\frac{4c_1}{3}\quad \mathrm{or}\quad c_4=0,~c_2=-2c_1,~c_3=0,
\end{equation}
which are consistent with case $\mathfrak{e}$ in Subsubsection \ref{III.2}. In both cases, $c_4=0$ holds, $i.e.$, the parity-odd term must vanish. Besides, after substituting the constraint given by the non-dynamical fields back into the quadratic   action, it will lead to $S^{(2)}_{v\left(W=0\right)}=0$. Therefore, the perturbations in vector sector with $W=0$ are not propagating.

\subsection{Quadratic Action for Scalar Perturbations}
Since $c_4=0$ is required in the vector  perturbations, the quadratic action for scalar perturbations in Fourier space when $W=0$ becomes 
\begin{equation}\label{saw0}
	S^{(2)}_{s\left(W=0\right)}=\int d\eta d^3\vec{k} ~\frac{a^2k^2}{2} \left\{2\left(2c_1-c_2\right)\lambda^{\prime2}-2\left(2c_1-c_2\right)k^2\lambda^2+k^2Z\left(B+\beta\right)^2-Z\left(A-\beta'\right)^2\right\}.
\end{equation}

It is clear that both $A$ and $B$ are not dynamical fields. Variation with respect to these fields yields the constraints:
\begin{equation}
	Z\left(B+\beta\right)=0,\quad Z\left(A-\beta'\right)=0,
\end{equation}
and whether $Z$ is equal to zero or not, the action (\ref{saw0}) will ultimately be simplified to
\begin{equation}
	S^{(2)}_{s(W=0)}=\int d\eta d^3\vec{k}~a^2k^2 \left(2c_1-c_2\right)\left(\lambda^{\prime2}-k^2\lambda^2\right).
\end{equation}

Therefore, it is necessary to require $2c_1-c_2\geq0$ to circumvent the occurrence of ghosts in the scalar perturbations. After combining the requirements  $2c_1+c_2\geq0$ in the tensor sector and (\ref{vw0}) in the vector sector, it can be concluded that there are two cases that are free of ghost instability when $W=0$. One is
\begin{equation}
	\begin{cases}
		c_2=2c_1\\
		c_3=-\frac{4c_1}{3}\\
		c_4=0\\
		c_1>0
	\end{cases},
\end{equation}
and in this case, vector and scalar perturbations are not dynamical, while the two tensor modes are both propagating. The other case is 
\begin{equation}
	\begin{cases}
		c_2=-2c_1\\
		c_3=c_4=0\\
		c_1>0
	\end{cases}.
\end{equation}
In this case, there will be only one scalar DoF remaining, while both the tensor and vector perturbations are not dynamical.

\section{Conclusion}\label{V}

In this paper, we studied an extended NGR model, which differs from the usual NGR model by the inclusion of additional parity-odd terms. These extra terms are quadratic in the torsion tensor and do not introduce any higher-order derivatives. After limiting to the quadratic form, totally there are only two independent parity-odd terms, one of which is the Nieh-Yan density, which is merely a total derivative and has no effect on the equation of motion if the coefficient is a constant. Therefore, only one parity-odd term needed to be taken into consideration in our model. Through investigations on the cosmological perturbations of this model, we obtained the results that this model can avoid the ghost instability in some regions of the parameter space, which is shown in Table.\ref{table}.
\begin{table}[h]
	\centering
	\caption{The range of parameters that can avoid the ghost instability}\label{table}
	\begin{minipage}{\textwidth}
		\begin{tabular*}{\textwidth}{c|c@{\extracolsep{\fill}}cc@{\extracolsep{\fill}}}
			\cline{1-4}
			{\bf Cases that free of ghost instability} &\multicolumn{3}{c}{{\bf Dynamical DoFs}}
			\\\cline{2-4}
			$~$ & \textbf{Tensor} & \textbf{Vector} & \textbf{Scalar} \\ 
			\cline{1-4}
			$c_4=0,~c_2=2c_1,~c_3=-4c_1~,c_1>0\quad$ & 2 & 0 & 0\\\cline{1-4}
			$c_4=0,~-2c_1<c_2<2c_1,~c_3=-2c_1-c_2,~c_1>0\quad$ & 2 & 0 & 1\\\cline{1-4}
			$c_4=0,~c_2=2c_1,~4c_1+c_3<0,~ c_1\geq0\quad$ &\quad 2 $(c_1>0)$ / 0 $(c_1=0)$ & 0 & 1\\\cline{1-4}
			$c_4=0,~c_2=-2c_1,~c_3<0,c_1>0\quad$ & 0 & 0 & 2\\\cline{1-4}
			$-\left|m\right|<c_4<\left|m\right|,~-2c_1<c_2<2c_1,~2c_1+c_2+c_3<0\quad$ & 2 & 4 & 2\\\cline{1-4}
			$c_4=0,~c_2=2c_1,~c_3=-4c_1/3,~c_1>0\quad$ & 2 & 0 & 0\\\cline{1-4}
			$c_4=0,~c_2=-2c_1,~c_3=0,~c_1>0\quad$ & 0 & 0 & 1\\\cline{1-4}		
		\end{tabular*}
	\end{minipage}
\end{table}
We found that in most cases of ghost free, the parity-odd term is required  to vanish. Nevertheless, one region in the parameter space is allowed in which the parity-odd term does exist, i.e., $c_4\neq 0$. In this case, all the scalar, vector, and tensor perturbations being are dynamical but healthy modes.

{\it Acknowledgement}: This work is supported by the National Key R\&D Program of China Grant No. 2021YFC2203102 and by NSFC under Grant No. 12075231 and 12247103.

{}
\end{document}